\documentclass[a4paper]{svproc}
\usepackage{url}

\usepackage{comment}
\usepackage{graphicx}
\usepackage{float}

\begin{document}

\mainmatter              

\title{An Online Development Environment for Answer Set Programming}

\titlerunning{An Online Development Environent}  

  \author{Elias Marcopoulos$^1$ and Christian Reotutar$^2$ and Yuanlin Zhang$^3$}

  \authorrunning{Marcopoulos, Reotutar and Zhang} 

%
%
%
%
  \institute{
    $^1$Department of Computer Science, Tufts University, USA \\
    $^2$Department of Computer Science, Johns Hopkins University, USA \\
    $^3$Texas Tech University, Lubbock, TX, USA \\
    \email{emarcopoulos@gmail.com, creotut1@jhu.edu, y.zhang@ttu.edu}
  }

\newcommand{\hide}[1]{}
\newcommand{\otherquestions}[1]{}
\newcommand{\catom}[2]{(\{#1\}, \{#2\})}
\newcommand{\set}[1]{\{#1\}}
\newcommand{\pg}[1]{{\tt #1}}
\newcommand{\emptyclause}{\Box}

\def\st{\medskip\noindent}

\newcommand{\kleene}{\wedge}

\newcommand{\ee}[1] {
  \begin{enumerate}
    #1 
  \end{enumerate}
}

\def\st{\medskip\noindent}

\maketitle              
\begin{abstract}
Recent progress in logic programming (e.g., the development
of the Answer Set Programming paradigm) has made it 
possible to teach it to general undergraduate and even 
high school students. 
Given the limited exposure of these students to
computer science, the complexity of downloading, installing
and using tools for writing logic programs could be a major 
barrier for logic programming to reach a much 
wider audience. 
We developed an online answer set programming environment with a self contained file system and a simple interface, 
allowing users to write logic programs and perform 
several tasks over the programs. 
\end{abstract}

\section{Introduction} 

\hide{Outline:
-- quick intro of state of the art of logic programming, its fitness to undergraduate and high school teaching [YL]
-- According to our teaching experience, a bottleneck in programming is the lack a "good" environment which has a low cost to maintain and is easy to learn and use

-- Our objective is to develop a user friendly environment 

-- Our solution is ... [following the logic in the poster] (emphasis online app, interface design, online folder and why we propose those. For folders, it is easy to navigate/share compared with a folder in Windows/Mac] etc.)
}
\hide{ Significant advances in theory and application of Declarative Programming have been made in the last two decades, thanks to the progress in both the language design and the solvers for constraint solving and logic based reasoning.
Examples include Constraint Programming \cite{rossi2006handbook} and Answer Set Programming \cite{GelK14} both of which originated from Artificial Intelligence studies.} 

Answer Set Programming (ASP) \cite{GelK14} is becoming a dominating language in the knowledge representation community 
\cite{McIlraith11,kowalski2014}
because it has offered elegant and effective solutions not 
only to classical Artificial Intelligence problems but 
also to many challenging application problems. Thanks to its 
simplicity and clarity in both informal and formal 
semantics, Answer Set Programming provides a ``natural"
modeling of many problems. At the same time, 
the fully declarative nature of ASP also cleared a 
major barrier to teach logic programming, as the procedural features of
classical logic programming systems such as PROLOG are
taken as the source of misconceptions in students' learning 
of Logic Programming \cite{mendelsohn1990programming}. 

ASP has been taught to undergraduate students, in the course of Artificial Intelligence, at Texas Tech for more
than a decade. We believe ASP has become mature enough
to be a language for us to introduce programming and problem solving to high school students. We have offered 
many sessions to 
students at New Deal High School and a three week long ASP course to high school students involved in the TexPREP program (http://www.math.ttu.edu/texprep/).  
In our teaching practice, we found that ASP is well accepted by the students and the students were able to focus on 
problem solving, instead of the language itself. The students were able to write programs to answer questions about the relationships (e.g., parent, ancestor) amongst family members and to find solutions for Sudoku problems. 

However, we have some 
major issues while using existing tools: installation of the tools 
to computers at a lab or at home is complex, and the existing tools are sensitive to the local settings of a computer. As a result, the flow of teaching the class 
was often interrupted by the problems associated with the use of the tools.   
Strong technical support needed for 
the management and use of the tools 
is prohibitive for teaching ASP to general undergraduate 
students or K-12 students.  
\hide{Since its installation is quite involved, the students were not able to 
install it on their own computer and thus lost the opportunity 
to practice the programming outside the class. 
}

During our teaching practice, we also found the need for
a more vivid presentation of the results of a logic program 
(more than just querying the program or getting the answer sets 
of the program). We also noted observations in literature that multimedia and 
visualization play a positive role in promoting students'
learning \cite{guzdial2001use,clark2009rethinking}.

To overcome the issues related to 
software tool management and use,
we have designed and built an online development 
environment for Answer Set Programming.  
The environment provides an editor for users 
to edit their programs, an online file system for 
them to store and retrieve their program and a few simple
buttons allows querying the program inside the
editor or getting answer 
sets of the program.  The environment 
uses SPARC \cite{BalaiGZ13} as the ASP language. 
SPARC is designed to further facilitate the teaching of logic programming by introducing sorts (or types) which simplify the difficult programming concept of {\em domain variables} in classical ASP systems such as Clingo \cite{gebser2011potassco} and help programmers to identify errors early thanks to sort information. Initial experiment of teaching SPARC to high school students is promising \cite{reyes2016using}.
To promote students' interests and learning, 
our environment also 
introduces  predicates 
for students to present their 
solutions to problems in
a more visually 
straightforward and 
exciting manner (instead of
the answer sets which are 
simply a set of literals). The URL for the online environment  
is \textit{http://goo.gl/ukSZET}.

The rest of the paper is organized as follows. 
Section~\ref{sec:sparc} recalls SPARC. The design 
and implementation of the online environment 
are presented in Section~\ref{sec:online}. The design 
and rendering of the drawing and animation predicates
are presented in Section~\ref{sec:drawing}. The paper
is concluded by Section~\ref{sec:discussion}.

\hide{
The language we are proposing can be used as an introduction to programming language, SPARC, is an example of an ASP programming language. According to \cite{dovierreasoning}, ASP has become a very promising course to be taught in a high school computer science class. This is because ASP teaches students about problem solving and using rigorous and precise definitions, according to \cite{reyes2016using}. 
}

\hide{
The only downside to ASP languages is that they produce an output of answer sets, which represent the knowledge of what is believed and not believed from a program by a rational thinker \cite{GelK14}. Answer sets are not always straightforward to interpret and understand. This can hurt a student's interest in learning computer science, but because of the drawings/animations offered through onlineSPARC we hypothesize that this possible setback may be overcome. Instead of having to interpret the answer sets, our system allows the display of answer sets in an intuitive way using drawings/animations! The creation of drawings/animations will not only peak the students' interests but will also serve to protect them from the deterrent of complex, unreadable answer sets.
}
\hide{ 
\section{Background on SPARC} 
SPARC is a logical programming language under answer set semantics which allows for the explicit representation of sorts. A SPARC program consists of three sections. The first section contains sort definitions and serves to define the sorts of objects in the program's domain. For example $\#color=\{blue, green\}$ is a sort definition. It means $\#color$ is a set of two colors, blue and green (as objects). The second section consists of a collection of predicate declarations in which the program's predicates and the sorts of their parameters are declared. $line\_color(\#style, \#color)$ is an example of a predicate declaration where $line\_color$ is a predicate and $\#style$ and $\#color$ are the sorts of parameters of $line_color$. Intuitively, $line\_color(X, Y)$ means that when drawing a line using style $X$, the color of the line is $Y$.  Finally, the rules section of SPARC consists of a collection of standard ASP rules. Consider the rule $$parent(X, Y) :- mother(X, Y).$$ 
where $parent(X, Y)$ means person $X$ is parent of $Y$ and $mother(X, Y)$ means person $X$ is the mother of $Y$, and $:-$ is read as ``if". This rule is read as for any $X$ and $Y$, $X$ is parent of $Y$ if $X$ is the mother of $Y$. 
}
\hide{
SPARC is a sorted ASP language that contains three sections:
\begin {itemize}
	\item sorts
	\begin{itemize}
    	\item Sorts are the objects of SPARC. 
        \item This is the section that the objects required for animation are declared.
        \item Some such objects include different drawing techniques, coordinates, and frames.
    \end{itemize}
	\item predicates
	\begin{itemize}
    	\item Predicates define relations between the sorts. 
       \item This is the section where the predicates draw and animate are declared.
    \end{itemize}
\item rules
	\begin{itemize}
    	\item Rules are the declarations of what is true and what is false. 
        \item Drawing and animation commands declared true will be animated.
    \end{itemize}
\end{itemize}

For example, the map coloring problem can easily be described using SPARC. The problem consists of coloring a map of the USA, where neighboring states can not have the same color. Consider three states: Texas, Oklahoma, and New Mexico. We can use {\tt neighbor(X,Y)} to denote that state {\tt X} is a neighbor of state {\tt Y}. We can represent Texas is a neighbor of New Mexico as {\tt neighbor(texas, newMexico).}  "We noted that $X$ is a neighbor of $Y$ if $Y$ is a neighbor $X$. This knowledge is represented as an ASP rule {\tt neighbor(X, Y) :- neighbor(Y, X).} where {\tt :-} is read as if. To express the intention that we would like to assign a color for {\tt r, g, b} to a state, we introduce a relation $colorOf(S, C)$ which denotes that the color of state $S$ is $C$. The knowledge that a color is assigned to a state is represented as an ASP rule {\tt color(S,r) | color(S,g) | color(S,b).} where {\tt | } is read as or. The rule reads, for any state S, S has a color of {\tt r}, {\tt g}, or {\tt b}. 
\\This structure was chosen as the background architecture for drawings and animations because "its simplicity and clarity in both informal and formal semantics ... provides natural modeling of many knowledge intensive problems" \cite{reyes2016using}. This hopefully will allow students learning computer science for the first time to have an easier time with creating these drawings and animations.
} 

\section{Answer Set Programming Language -- SPARC}
\label{sec:sparc}
\hide{
-- Introduction of ASP and demo ease of programming [YL]
}

SPARC is an Answer Set Programming language  which allows for the explicit representation of sorts. A SPARC program consists of three sections: {\em sorts}, {\em predicates} and {\em rules}.  
We will use the map coloring problem as an example to illustrate
SPARC: can the USA map be colored using red, green and blue 
such that no two neighboring states have  the same color? 

The first step is to identify the objects and their sorts in the problem. 
For example, the three colors are important and they form the 
sort of color for this problem. In SPARC syntax, we use 
$\#color=\{red, green, blue\}$ to represent the 
objects and their sort.
The sorts section of the SPARC program is 

\begin{verbatim}
sorts % the keyword to start the sorts section
#color = {red,green,blue}.
#state = {texas, colorado, newMexico, ......}. 
\end{verbatim}

The next step is to identify relations in the problem and declare in
the predicates section the sorts of the parameters of the predicates
corresponding to the relations. The predicates section of the program is 

\begin{verbatim}
predicates % the keyword to start the predicates section
  % neighbor(X, Y) denotes that state X is a neighbor of state Y. 
  neighbor(#state, #state). 
  % ofColor(X, C) denotes that state X has color C
  ofColor(#state, #color). 
\end{verbatim}

The last step is to identify the knowledge needed in the problem 
and translate it into rules. The rules section of a SPARC program 
consists of rules in the typical ASP syntax. 
The rules section of a SPARC program will include the following.  

\begin{verbatim}
rules  % the keyword to start the rules section
  % Texas is a neighor of Colorado 
  neighbor(texas, colorado).
  % The neighbor relation is symetric
  neighbor(S1, S2) :- neighbor(S2, S1). 
  % Any state has one of the three colors: red, green and blue
  ofColor(S, red) | ofColor(S, green) | ofColor(S, blue). 
  % No two neighbors have the same color 
  :- ofColor(S1, C), ofColor(S2, C),  neighbor(S1, S2), S1 != S2. 
\end{verbatim}

\hide{-- Introduction of existing development environment (a quick survey of the state of the art dev env can start by reading \cite{FebbraroRR11}, and difference beteween our work and the existing work. 
}

\hide{
These tools help a lot in our teaching but still present enough challenge. As an expert, when we teach 
ASP to New Deal High School students, we have an IT support member from our university who worked with
the IT support personnel in the high school to install the ASPIDE in their lab. 

But the ASPIDE on a majority of computers failed to work in our last hands-on session (no hands-on in the previous sessions). Similarly in our summer school for TexPREP students. Although we had a very experienced IT professional to install and manage ASPIDE, there was frequent long interrupt almost for each class 
due to the failure of ASPIDE on some of the computers, which negatively affected the flow of the teaching. }
\section{Online Development Environment Design and Implementation} 
\label{sec:online}

\subsection{Environment Design}

The principle of the design is that the environment,
with the simplest possible interface, 
should provide full support, from writing programming to 
getting the answer sets of the program, for teaching 
Answer Set Programming.  

\begin{figure}[H]

	\centering
	\includegraphics[width=1\textwidth]{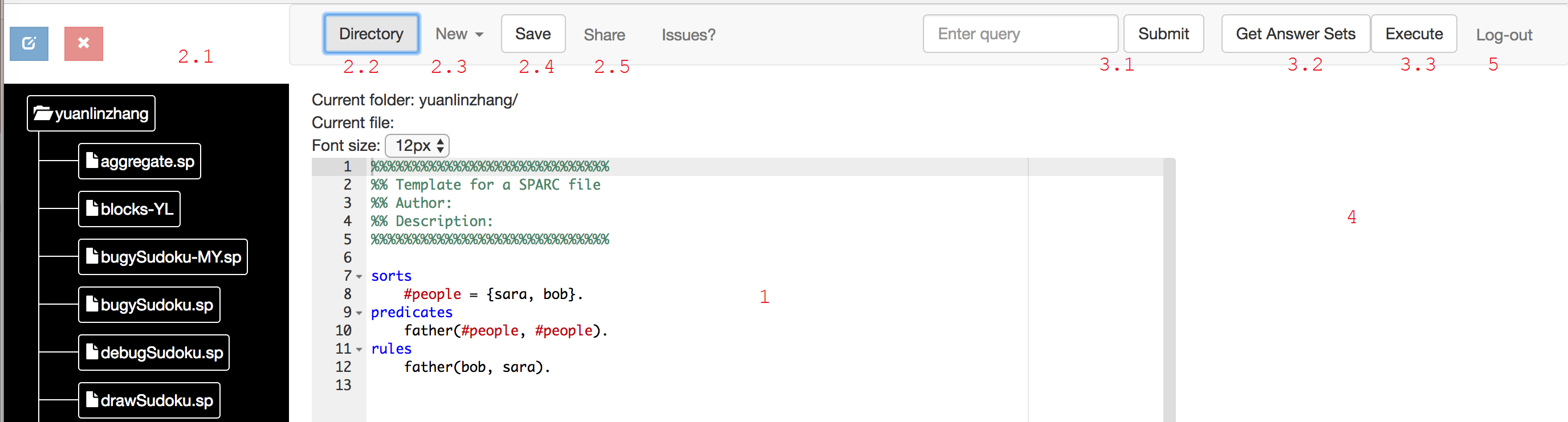}
	\caption{\label{fig:onlineSPARC} User Interface of the System (the red numbers indicate the areas/components in the interface)}
	
\end{figure}

The design of the interface is shown in Figure~\ref{fig:onlineSPARC}. It
consists of 3 components: 1) the editor to edit a program, 2) the file navigation 
system and 3) the operations over the program.

One can edit a SPARC program directly inside the editor which 
has syntax highlighting features (area 1).  The file inside the editor 
can be saved by clicking the ``Save" button (2.4). The files and folders 
are displayed in the area 2.1.  The user can traverse them using the mouse 
like traversing a file system on a typical operating system.  
Files can be deleted and their names can be changed. To create a folder or
a file, one clicks the ``New" button (2.3). The panel showing files/folders can be toggled 
by clicking the ``Directory" button (2.2) (so that users can have more space
for the editing or result area (4)). 
To ask queries to the program inside the editor, one can type a query (a conjunction of literals) in the text box (3.1) and then press the ``Submit" 
button (3.1).
The answer to the query will be shown in area 4. 
For a ground query (i.e., a query without 
variables), the answer is {\em yes} if every literal in the query is in every 
answer set of the program,  is {\em no} if the complement ($p$ and $\neg p$, where $p$ is an atom, are complements) of some literal is in every answer set
of the program, and {\em unknown} otherwise. An answer to a query with variables is a set of ground terms for the variables in the query such that 
the answer to the query resulting from replacing the variables by the corresponding ground terms is yes. Formal definitions of queries and answers to queries can be found in Section~2.2 of \cite{GelK14}.
To see the answer sets of a program, click  the  ``Get Answer Sets" button (3.2).  
When ``Execute" button (3.3) is clicked, the atoms with drawing and animation 
in the answer set of the program will be rendered in the display area (4). (For now, 
when there is more than one answer set, the environment displays an error.)

A user can only access the full interface discussed above 
after login. The user will log out by clicking the ``Logout" button (5). 
Without login,  the interface is much simpler, with all the file navigation
related functionalities invisible.  Such an interface is convenient
for testing or doing a quick demo of a SPARC program.   

\subsection{Implementation}. 

The architecture of the online environment follows 
that of a typical web application. Is consists of a front
end component and a back end component. 
The front end provides the user interface and sends 
users' request to the back end, and the back end 
fulfills the request and returns results, if needed, 
back to the front end. After getting the results
from the back end, the front end will update the 
interface correspondingly (e.g., display query 
answers to the result area). Details about the components
and their interactions are given below. 

\st {\bf Front End}. The front end is implemented by HTML and 
JavaScript. The editor in our front end uses  
ACE which is an embeddable (to any web page) code 
editor written in JavaScript ({\tt https://ace.c9.io/}). 
The panel for file/folder navigation is based on JavaScript 
code by Yuez.me. 

\st {\bf Back End and Interactions between the Front End and the Back End}.  The back end is mainly implemented using PHP and is hosted 
on the server side. It has three components: 
1) file system management, 2) inference engine and 3) drawing/animation rendering. 

The {\bf file system management} 
uses a database to manage the files
and folders of all users of the environment. The ER diagram of the
system is shown below:
\begin{figure}[H]
	\begin{center}
		\includegraphics[width=0.7\textwidth]{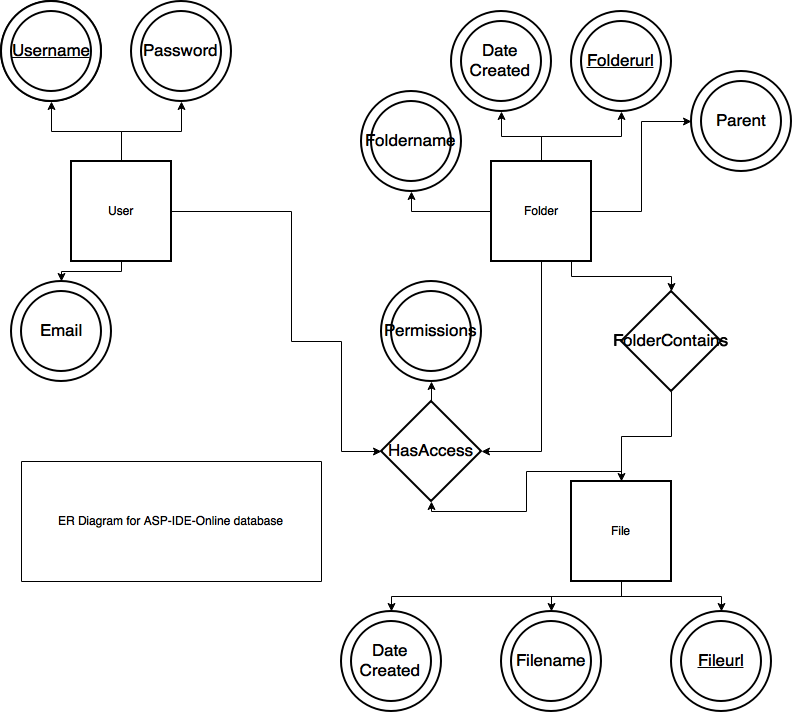}
	\end{center}
	\caption{The ER diagram for file/folder management. Most names have a straightforward meaning. The Folderurl and Fileurl above 
	refer to the full path of the folder/file in the file system.}
\end{figure}

The SPARC files are saved in the server file system, not in a database table.  The sharing is managed by the sharing information in the relevant database tables. 
In our implementation, we use mySQL database 
system. 

The file management system gets request such as creating a new file/folder, deleting a file, saving a file, getting the files and folders, etc, 
from the front end. It then updates the tables and local file 
system correspondingly and returns the needed results to 
the front end. After the front end gets the results, it will 
update the graphical user interface (e.g., display 
the program returned from the back end inside the 
editor) if needed. 

The {\bf inference engine} gets the request of answering a query or obtaining all answer sets of a program. It calls the SPARC solver 
\cite{BalaiGZ13} to find all answer sets. Then in terms of these
answer sets, it returns requested information to the front end. 
After the front end gets the response from the back end, it will 
show the result in the display area of the web page. 

Details of the design and implementation of
 {\bf drawing/animation rendering}  can be found in Section~\ref{sec:drawingImplementation}.

\section{Drawing and Animation Design and Implementation}
\label{sec:drawing}
\hide{
\begin{verbatim}
Elias: you will work on this section. 
Related document: 
 1. on shared google drive: onlineSPARCviz-2016REU
  animation manual/
  animationReport.docx
 2. Different versions of earlier paper EAAI
    https://www.overleaf.com/5712007swnvfm#/18622321/ 
 3. Some comments, in the current version, left from 
    earlier version. 
  
Todo: 
  audience: researchers know Logic Programming and even 
            ASP pretty well. 
  Extend Section 3.1 as needded (e.g., in terms of your manual 
    and report on google drive). In the section, only 
    include information a programmer needs to know to 
    do drawing and animation. 3.1 is currently in a 
    decent shape. 
  Section 3.2 needs more substantial extension. You can
    include the algorithm [in fact, we do have an algorithm 
    in this version, but commented away using \hide] 
    and other implementation information 
    (e.g., HTML Canvas 5 etc.). 
  pages: you have 2.5 *additional* pages for Section 3.
    You don't have to use up all of them. If you need 
    more pages, let me know as early as you can. 
\end{verbatim}
}
\subsection{Drawing and Animation Design}
\hide{ We may use the following format. (This is just a writing outline which will be deleted in the final version.)
\begin{itemize}
  \item intro 
  \item syntax for drawing and intuitive meaning
  \item Examples to further illustrate the meaning and to illustrate its use. 
  \item Syntax for animation (some basics of animation may be needed before or after 
  \item Examples for animation ...
  
\end{itemize}
}

To allow programmers to create drawings and animations using SPARC, we simply design two predicates, called {\em display predicates}: one for drawing and one for animation. 
The atoms using these predicates are called {\em display atoms}. To use these atoms
in a SPARC program, a programmer 
needs to include 
sorts (e.g., sort of colors, 
fonts and numbers) and the corresponding predicate declaration which 
are predefined. In the following,
we only focus on the atoms and their 
use for drawing and animation. 

\medskip
\hide{
\noindent {\bf Styling}
Before we can think about displaying drawings or animations, we must first define some terms to do with styling. For example, we may introduce a style name 
{\tt greenline} and associate it to the green color by the {\em style command} {\tt line\_color(greenline, green)}.
In general, there are two variations of style commands, ones that modify text style and ones that modify line style.
\begin{itemize}
\item could talk more in detail about styling here
\item add a section about shape
\item expand on what the canvas is
\item two types of styling
\item more about stylenames
\item maybe include more commands in general in this section that can be executed?
\item drawing
\item subsection shape
\item subsection style
\end{itemize}
}
\noindent {\bf Drawing}. 
A {\em drawing predicate} is of the form: {\tt draw($c$)}
where $c$ is called a {\em drawing command}.
Intuitively the atom containing this predicate draws texts and graphics as instructed by the command $c$. By drawing a picture, we mean a {\em shape} is drawn with a {\em style}.  We define
a {\em shape} as either text or a geometric line or curve. Also, a {\em style} specifies the physical visual properties of the shape it is 
applied to. For example, visual properties include color, thickness, and font. For modularity, we introduce {\em style names}, which are labels that can be associated with different styles so that the style may be reused without being redefined. A drawing is completed by associating this shape and style to a certain position in the {\em canvas}, which is simply the display board. Note, the origin of the coordinate system is at the top left corner of the canvas.

Here is a an example of drawing a red line from point $(0,0)$ to  $(2,2)$. First, we introduce a style name 
{\tt redline} and associate it to the red color by the {\em style command} {\tt line\_color(redline, red)}. With this defined style we then draw the red line by the {\em shape command} {\tt draw\_line(redline, 0, 0, 2, 2)}. Style commands and shape commands form all drawing commands.
The SPARC program rules to draw the given line are 

{\tt draw(line\_color(redline, red))}.

{\tt draw(draw\_line(redline, 0, 0, 2, 2))}.

The style commands of our system include the following:\\
{\tt linewidth(sn, t)} specifies that lines drawn with style name {\tt sn} should be drawn with a line thickness {\tt t}.
{\tt textfont(sn, fs, ff)} specifies that text drawn with style name {\tt sn} should be drawn with a font size {\tt fs} and a font family {\tt ff}.
{\tt linecap(sn, c)} specifies that lines drawn with style name {\tt sn} should be drawn with a capping {\tt c}, such as an arrowhead.
{\tt textalign(sn, al)} specifies that text drawn with style name {\tt sn} should be drawn with an alignment on the page {\tt al}.
{\tt line\_color(sn, c)} specifies that lines drawn with style name {\tt sn} should be drawn with a color {\tt c}.
{\tt textcolor(sn, c)} specifies that text drawn with style name {\tt sn} should be drawn with a color {\tt c}.

The shape commands include the following:\\ 
{\tt draw\_line(sn, xs, ys, xe, ye)} draws a line
from starting point {\tt (xs, ys)} to ending point 
{\tt (xe, ye)} with style name {\tt sn}; 
{\tt draw\_quad\_curve(sn, xs, ys, bx, by, xe, ye)}
draws a quadratic Bezier curve, with style name
{\tt sn}, from the current point {\tt (xs, ys)} 
to the end point {\tt (xe, ye)} using the control point {\tt (bx, by)}; 
{\tt draw\_bezier\_curve(sn, xs, ys, b1x, b1y, b2x, b2y, xe, ye)} draws a cubic Bezier curve, using
style name {\tt sn}, from the current point {\tt (xs, ys)} to the end point {\tt (xe, ye)} using the control points {\tt (b1x, b1y)} and {\tt (b2x, b2y)}; 
{\tt draw\_arc\_curve(sn, xs, ys, r, sa, se)} draws 
an arc using style name {\tt sn} and the arc is centered at 
{\tt (x, y)} with radius {\tt r} starting at angle {\tt sa} and ending at angle {\tt se} going in the clockwise direction;
{\tt draw\_text(sn, x, xs, ys)} prints value of {\tt x} as text
to screen from point {\tt (xs, ys)} using 
style name {\tt sn}.

\medskip \noindent {\bf Animation}.
A {\em frame}, a basic concept in animation, is defined as a drawing.
When a sequence of frames, whose content is normally relevant, is shown on the screen in rapid succession (usually 24, 25, 30, or 60 frames per second), a fluid animation is seemingly created. To design an animation, a designer 
will specify the drawing for each frame. Given that the order of frames matters, we give a frame a value equal to its index in a sequence of frames. We introduce the {\em animate predicate} { \tt animate($c, i$)}
which indicates a desire to draw a picture at the $i^{th}$ frame using drawing command $c$ and $i$ starts from $0$. 
The frames will be shown on the screen at a rate of 60 frames per second, and the $i^{th}$ frame will be showed at time $(i*1/60)$ 
(in a unit of second) from the start of the animation for a duration of $1/60$ of a second.

As an example, we would like to elaborate on an animation where a red box (with side length of 10 pixels) moves from the point $(1,70)$ to $(200, 70)$. 
We will create 200 frames with the box (whose bottom left corner is) at point $(i+1, 70)$ in $i^{th}$ frame.

Let the variable I be of a sort called frame, defined from 0 to some large number.
In every frame $I$, we specify the drawing styling $redline$: 

\noindent {\tt animate(line\_color(redline, red), I).}

To make a box at the $I^{th}$ frame, we need to draw the box's four sides using the style associated with style name {\tt redline}. The following describes the four sides of a box at any frame: bottom - $(I+1,70)$ to $(I+1+10, 70)$, left - $(I+1, 70)$ to $(I+1, 60)$, top - $(I+1, 60)$ to $(I+1+10, 60)$ and right - $(I+1+10, 60)$ to $(I+1+10, 70)$. Hence we have the rules 

\noindent {\tt \small animate(draw\_line(redline,I+1,70,I+11,70),I).}

\noindent {\tt \small animate(draw\_line(redline,I+1,70,I+1,60),I).}

\noindent {\tt \small animate(draw\_line(redline,I+1,60,I+11,60),I).}

\noindent {\tt \small animate(draw\_line(redline,I+11,60,I+11,70),I).}

Note that the drawing predicate produces the intended drawing throughout all the frames 
creating a static drawing. On the other hand, the animate predicate produces a drawing only for a specific frame.


\subsection{Algorithm and Implementation}
\label{sec:drawingImplementation}

\hide{
The input to the main algorithm is a SPARC program $P$. The output is an HTML5 program containing a canvas which will be rendered by the browser. The algorithm finds an answer set (i.e., all atoms that are true under the program by stable model semantics \cite{GelK14}), extracts all display atoms, and generates an HTML5 program that uses canvas to set the drawing style properly according to the 
style atoms for the $i^{th}$ frame and then renders all shape commands specified by the animate atoms for the $i^{th}$ frame. The drawing commands inside the display atoms will be rendered for every frame. (An optimization is made to reduce repeated rendering efforts.)}

We first define our input and output: The input to the main algorithm is a 
SPARC program $P$. The output is an HTML5 program segment containing a canvas 
element which will be rendered by an Internet browser. 
A key part of our 
algorithm is to render the display atoms (specified in the answer set of $P$) using canvas methods. 

HTML5 canvas element is used to draw graphics via scripting using JavaScript. In the following, we will use an example to demonstrate how
a drawing command is implemented by JavaScript code using canvas methods.
Consider again

{\tt draw(line\_color(redline, red))}.

{\tt draw(draw\_line(redline, 0, 0, 2, 2))}.

When we render the shape command {\tt draw\_line}, we need 
to know the meaning of the {\tt redline} style. 
From the style command {\tt line\_color}, 
we know it means {\tt red}. 
We first create an object {\tt ctx} for a given
canvas (simply identified by a name) where
we would like to render the display atoms. 
The object offers methods to render the graphics in the canvas. We then use the following JavaScript code to implement the
shape command to draw a line from (0,0) to (2,2): \\
{\tt ctx.beginPath();}\\
{\tt ctx.moveTo(0,0);}\\
{\tt ctx.lineTo(2,2);}\\
{\tt ctx.stroke();}

To make the line in red color, we have to insert the following 
JavaScript statement before the {\tt ctx.stroke()} in
the code above: \\
{\tt ctx.strokeStyle="red";} 

The meaning of the canvas methods in the code above is straightforward. 
We don't explain them further. 
Now we are in a position to present the algorithm. 

\hide{ [??Elias, clarify the following sentences??] However, all of this JavaScript code is abstracted away, and is only reproduces here to help the reader understand what is accomplished in the following full delineated algorithm for drawing and animation:}


\st Algorithm:
\begin{itemize}
\item Input: a SPARC program $P$ with display predicates.

\item Output: a HTML program segment which allows the rendering of the display atoms in the answer set of $P$ in an Internet Browser.

\item Steps:
\begin{enumerate}
 	\item Call SPARC solver to obtain an answer set $S$ of $P$.
 
 \item Let  $script$ be an array of empty strings. $script[i]$ will hold the JavaScript statements to render the graphics for $i^{th}$ frame.
 \item For each display atom $a$ in $S$, 
 \begin{itemize}
	\item If any error is found in the display atoms, present an error to the user detailing the incorrect usage of the atoms. 	
    \item If $a$ contains a shape command, let its style name be $sn$, find all style commands defining $sn$. For each style command, translate it into the corresponding JavaScript code $P_s$ on modifying the styling of the canvas pen. Then translate the shape command into JavaScript code $P_r$ that renders that command. Let $P_d$ be the proper combination of $P_s$ and $P_r$ to 
    render $a$. 
    \begin{itemize}
      \item if $a$ is an drawing atom, append $P_d$  
      to $script[i]$ for every frame $i$ of the animation. 
      \item if $a$ is an animation atom, let $i$ be the frame referred to in $a$. Append $P_d$ to $script[i]$. 
    \end{itemize}
 \end{itemize}
 \item Formulate the output program $P$ as follows:
 \begin{itemize}
  \item add, to $P$, the canvas element {\tt <canvas id="myCanvas" width="500" height="500"> </canvas>}. 
  \item add, to $P$, the script element {\tt <script> </script>} whose content includes 
	\begin{itemize} 
	  \item the JavaScript code to associate the drawings in this script element with the canvas element above.
	  \item an array $drawings$ initialized by the content of $script$ array.
	  \item Javascript code executing the statements in $drawings[i]$ when the time to show frame $i$ starts.
	\end{itemize}
  \end{itemize}
 
 \hide{
 \item Print the created array into a JavaScript tag within an HTML program along with all the necessary code to start the animation when it is loaded by a browser:	
 \begin{itemize}
 \item print code to create the HTML5 canvas element in the browser
 \item for each array of commands corresponding with a frame
 \begin{itemize}
 \item print out the styling commands associated with the frame
 \item print out the drawing commands associated with the frame
 \end{itemize}
 \item print code to execute each frame consecutively whenever the browser refreshes using requestAnimationFrame JavaScript function
 \item set off first frame by initial call to requestAnimationFrame
 \end{itemize}}
 \end{enumerate}
\end{itemize} 
\noindent End of algorithm. 

\st {\bf Implementation}. The ``Execute" button in the webpage (front end)  of the online SPARC  environment is for programmers to render the display atoms in the answer set of their programs. 
The Java program implementing our algorithm 
above is at
the server side. When the ``Execute" button is clicked,
the programmer's SPARC program will be 
sent to the server side and the algorithm will
be invoked with the program. The output
(i.e., the canvas 
and script elements)  of the 
algorithm will be sent back to the front end 
and the JavaScript in the front end will catch 
the output and insert it into the result
display area of the front web page (See Figure~\ref{fig:onlineSPARC}). The Internet
browser will then automatically render the 
updated web page and the drawing or animation
will be rendered as a result. 

\hide{
	The web page of our environment is shown in Figure 1.

\begin{figure}[H]
\begin{center}
\includegraphics[width=0.7\textwidth]{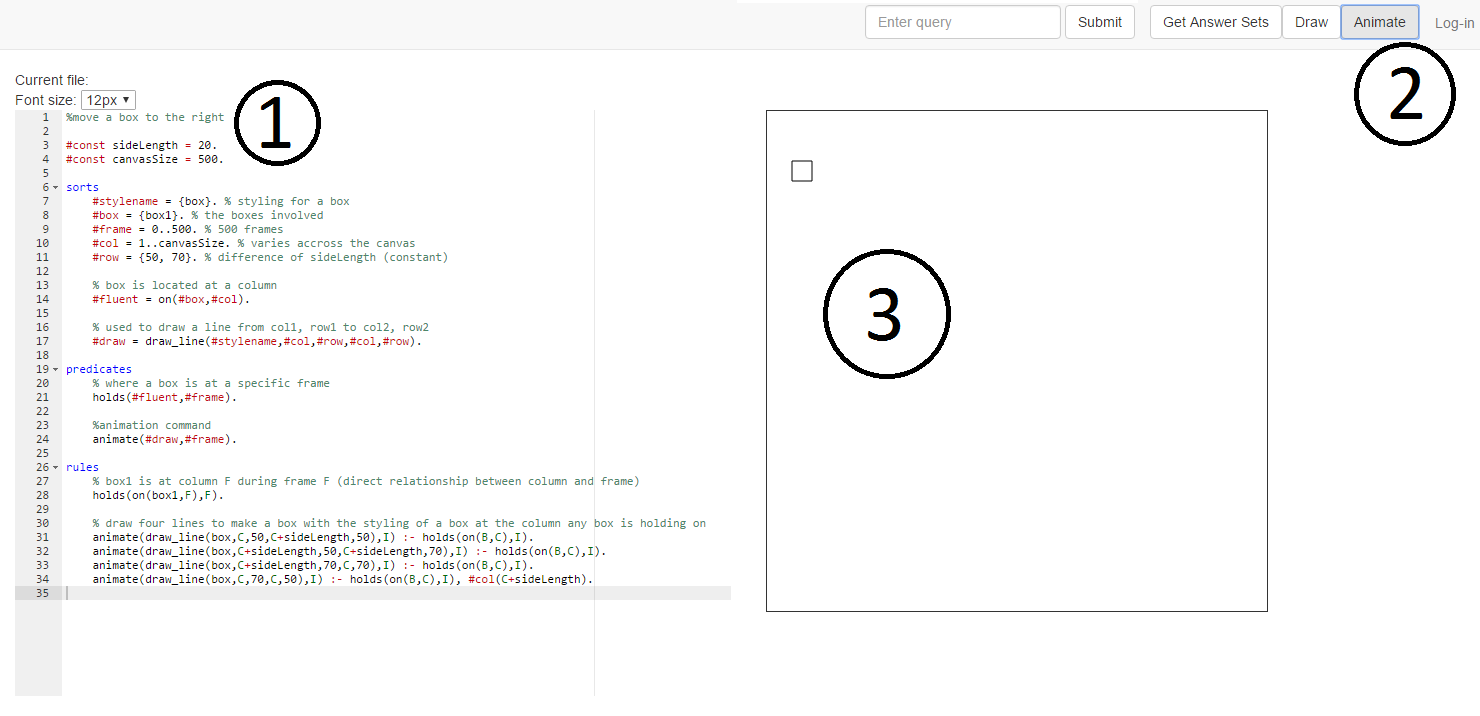}
\end{center}
\caption{Online SPARC Environment (The programmer can edit a SPARC program in area 1. After clicking the Execute button (area 2), the drawing/animation will be shown in area 3.)}
\end{figure}
}

Example SPARC programs with drawing and animation can be found at https://goo.gl/nLD4LD. 

\hide{ output in the 
 The script element from the algorithm will be inserted immediately below the canvas element in the main PHP program for our on-line system, and the browser will render the drawing and animation inside the canvas element when the script is automatically run.}

\hide{

We integrated our algorithm into the existing online SPARC environment. The interface of our environment is shown in Figure 1.
\begin{figure}[H]
\includegraphics[width=0.45\textwidth]{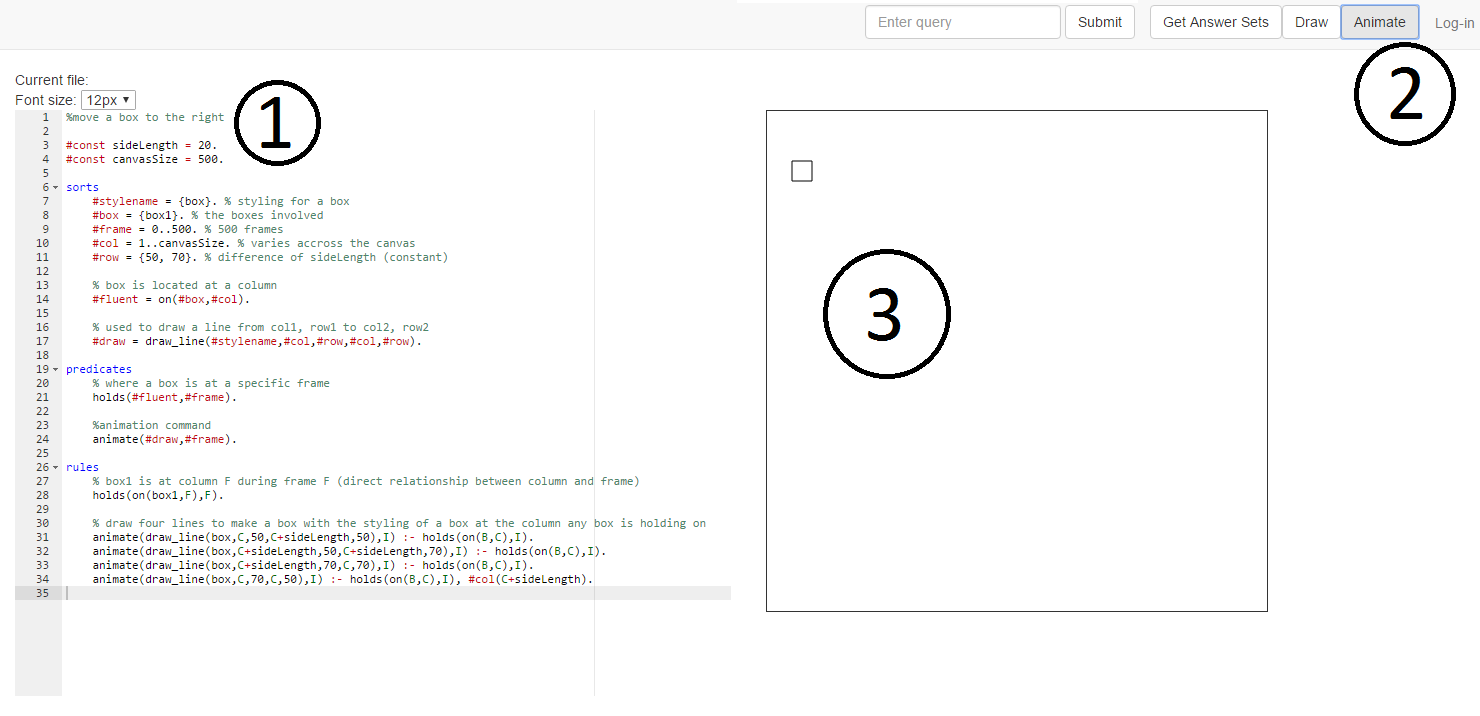}
\caption{Online SPARC Environment}
\end{figure}

}

\hide{ 
The number labels above correspond with the numbered list below:

\begin{enumerate}
\item Writing an Animation Program 
\begin{itemize}
\item One simply types the code into the box that appears at the left side of the web page (Figure 1.1). It is important that these programs contain sorts for different drawing commands and also either the animation command predicate 'animate' or 'draw', as those are how animations will be executed.
\end{itemize}
\item Executing an Animation Program
\begin{itemize}
\item To execute an animation all one needs to do is click on the button "Animate" in the top right corner of the web page (Figure 1.2). The code will be sent to a solver, and then the answer set of the program will be parsed to produce a JavaScript animation using HTML Canvas. 
\end{itemize}
\item Watching an Animation
\begin{itemize}
\item After clicking the "Animate" button an animation will be produced in the canvas that is on the right of the web page(Figure 1.3). If no animation is produced make sure the animation and drawing commands are properly defined. 
\end{itemize}
\end{enumerate}
}

\section{Discussion and Related Work}
\label{sec:discussion}

\hide{ YL: by discussion I mean both related work and future work. I didn't revise your writing, but just insert a piece to show a flavor of writing a comparison with existing work. }

\hide{
By studying the issues in our teaching practices, we realized the following obstacles.1)  The existing tools are standalone software and it is expensive to maintain those tools during and outside of class. 
2) The tools use of the local computer environment needs students to have some knowledge of directory structures and how they are connected to the development tools. 
3) The complex user interfaces are packed with many functions that distract the attention of students from the key ASP concepts and from problem solving. 
4) Sharing created programs (e.g., submission to the instructors) is challenging. }

As ASP has been applied to more and more problems,  the 
importance of ASP software development tools has been 
realized by the community.  
Some integrated development environment (IDE) tools, e.g., APE \cite{sureshkumar2007ape}, ASPIDE\cite{FebbraroRR11}, iGROM\cite{iGROM} and SeaLion \cite{oetsch2013sealion} 
have previously been developed.  
They provide a graphical user interface for users to carry out a sequence 
of tasks from editing an ASP program to debugging that program, easing 
the use of ASP significantly. However, the target audience of these tools 
is experienced software developers. Compared with the existing environments, our environment is online, self contained (i.e., fully 
independent of the users' local computers) and 
provides a very simple interface,  focusing on teaching only. 
The interface is operable by any person who is able to use a typical
web site and traverse a local file system.  

As for drawing and animation, our work is based on 
the work of Cliffe et al.
\cite{cliffe2008aspviz}. They are the first to introduce, to ASP,  a design of display predicates and to render drawings and animations using the program ASPviz. Our drawing commands are similar to theirs. 
The syntax of their animation atoms is not clear from their paper. It seems 
(from examples on github at {\tt goo.gl/kgUzJK} accessed on 4/30/2017)
that multiple answer sets may be needed to produce an animation. 
In our work we use a design where the programmers
are allowed to draw at any frame (specifying a range of the frames) and 
the real time difference between two neighboring frames is 1/60 second.
Another clear difference is that our implementation is online while 
theirs is a standalone software.  A more recent 
system, Kara, a standalone software  by Kloimullner et al. \cite{kloimullner2013kara}, deals with drawing only. Another system
ARVis \cite{ambroz2013arvis} offers method 
to visualize the relations between answer sets
of a given program. 
We also note an online environment for IDP (which is a knowledge representation paradigm close to ASP)  by Dasseville and Janssens \cite{dasseville2015web}. 
It also utilizes a very simple interface for the IDP system and  allows 
drawing and animation using IDP through IDPD3 (a library to visualize
models of logic theories) by Lapauw et al. \cite{lapauw2015visualising}.
In addition to drawing and animation, IDPD3 allows users' interaction 
with the IDP program (although in a limited manner in its current
implementation), which is absent from most other systems including 
ours.
Our environment is also different from the online IDP environment in
that ours targets ASP and offers an online file system. 
Both DLV and Clingo offer online environments ({\tt http://asptut.gibbi.com/} and
{\tt http://potassco.sourceforge.net/clingo.html} respectively) 
which provide an editor and a window to show the output of 
the execution of dlv and clingo command, but provide no other functionalities. We also noted the SWISH \\ ({\tt http://lpsdemo.interprolog.com}) which offers 
an online environment for Prolog and a more
recent computer language Logic-based Production 
Systems \cite{kowalski2016programming}. 
A unique functionality of our online environment is to query a program.  
It allows to teach (particular to general students) 
basics of Logic Programming without first 
touching the full concept of answer sets. 

\hide{
Most existing programming environments are proved useful for experienced programmers, but it is still challenging to use them in teaching students who are novices in programming. Our online environment, with a carefully designed simple interface and a self contained file system,  provided easy access and sharing, reduced the learning curve, and removed the installation and maintenance expenses, by our experience of using it in our teaching in Fall 2015. We take it as an enabler to teach ASP to more students, including students in high school. }

\hide{ 
However, the animate predicate is different from theirs (?? Can someone make the difference very specific in the design section?)
and we feel that our design is more natural in modeling animation tasks ( future work is to study these two designs -- goes to future work part). There are a few other major differences between our work and theirs. We use SPARC (e.g., sorts etc make it easier/ more effective to write correct programs compared with Clingo \cite{gebser2011potassco} or DLV cite...). Next, ASPviz requires two programs one for drawing/animation while the other 
is for information needed by the drawing/animation. We don't 
require them to be separate program although we encourage the methodology of organize one program into two parts. 
Finally, we provide an online environment instead of a standalone application with the advantage: very expensive to install and manage a standalone program in a teaching environment for genera students particularly younger students (e.g., high school ones) \cite{reyes2016using}). Our online version allows us to HTML canvas will significantly reduce the development efforts. }

\hide{ 
Some of the challenges that when overcome will improve our online environment is organizing the server in a more user friendly and functional manner. Also, improving the efficiency of the solvers to be able to produce more complicated animations in less time, so that more time can be spent learning and coding.} 

\hide{It is noted that thanks to ASP/SPARC rules, 
one can define more abstract and easy to use
drawing/animation ``commands'' from the base set of commands.
With the new drawing and animation features,
students can not only solve problems such as Sudoku and AI problems, but can also present the results in vivid and straightforward drawings and animations. We hope 
the new environment will inspire more interest in Logic Programming, AI, and computer science in general, as well as provide a more effective learning environment. }
\hide{
and classical AI planning problems but also 
more students may be inspired to learn computer science. W
e hope that schools will see the value in ASP and begin to use it with the aid of the animation tools.}
When we outreached to a
local high school before, we needed an experienced student 
to communicate with the school lab several times before the 
final installation of the software on their computers could be completed. A carefully 
drafted document is prepared for students to install the software 
on their computers. There are still unexpected issues 
during lab or when students use/install the software at home. 
These difficulties made it almost impossible to outreach to
the high school with success. 
With the availability of our online environment, we only 
need to focus on the teaching content of ASP without 
worrying about the technical support.  We hope our environment, and other online environments, for knowledge representation systems will 
expand the teaching of knowledge representation to 
a much wider audience in the future. The drawing 
and animation are new features of the online 
environment and was not tested in high school 
teaching. We have used the drawing and animation
in a senior year course -- special topics in AI -- 
in spring 2017. 
Students demonstrated interests in drawing and animation
and they were able to produce interesting 
animation. We also noted that it can be  
very slow for ASP solvers to produce 
the answer set of an animation program 
when the ground program is big. 

In the future, it will be interesting to have a 
more rigorous evaluation of the online environment.

\section{Acknowledgments}
The authors were partially supported by National Science Foundation (grant\# CNS-1359359). 
We thank Evgenii Balai, Mbathio Diagne, Michael Degraw, Peter Lee, Maede Rayatidamavandi,  Crisel Suarez, Edward Wertz and Shao-Lon Yeh
for their contribution to the implementation of the environment. 
We thank Michael Gelfond and Yinan Zhang for their input and help.  

\bibliographystyle{splncs03}
\bibliography{biblio,onlineSPARC,onlineSPARC-d}
\end{document}